\documentclass{article}
\usepackage{graphicx} 
\usepackage{amsmath}
\usepackage{color}
\usepackage{hyperref}
\usepackage{authblk}
\begin{document}
\title{The effect of Coulomb interaction on Lévy source parameters}

\author[1,2]{Kristóf Csiki}
\author[2]{Máté Csanád}
\author[2,3]{Barnabás Pórfy}
\affil[1]{Vörösmarty Mihály High School, Budapest, Hungary}
\affil[2]{Department of Atomic Physics, ELTE E\"otv\"os Lor\'and University, Budapest, Hungary}
\affil[3]{HUN-REN Wigner Research Centre for Physics, Budapest, Hungary}
\maketitle

\begin{abstract}
Femtoscopy plays an important role in the study of heavy-ion collisions and particle sources generated within them. In both experimental measurements and theoretical models, a Lévy distribution is used to describe the shape of the source, as the Gaussian distribution has proven to be inaccurate based on experimental observations. Despite this, there are still discrepancies between experimental data and the results of theoretical models. One reason may be that these theoretical models either approximate the Coulomb interaction poorly or omit it entirely. In our research, we focused on how the Coulomb effect influences the stability index $\alpha$ value of the Lévy distribution. For this, we used a custom-developed simulation framework written in C++ and Python. The C++ module integrates the equations of motion for relativistic pions using the Leapfrog algorithm, while the Python post-processing script generates the two-particle spatial source function from the phase-space coordinates. We investigate the ratio and slope of the generated and pair-source fitted $\alpha$ as a function of the number of particles, fitting range, and the initial source size and density.

\end{abstract}

\section{Introduction}

High-energy heavy-ion collisions produce the quark--gluon plasma (QGP), and understanding its properties is one of the main goals of many experiments. Femtoscopy, the study of quantum-statistical correlations between particles~\cite{Goldhaber:1959mj, Goldhaber:1960sf}, is a key tool in uncovering certain properties, as it gives access to the femtometer-scale structure of the particle-emitting source~\cite{Lisa:2005dd, Csorgo:1999sj}. For some time, the shape of particle emission was assumed to be Gaussian~\cite{Csorgo:1994fg, Akkelin:1995gh}. Measurements in various systems at several different energy scales energies~\cite{PHENIX:2017ino,NA61SHINE:2023qzr,CMS:2023xyd,Kincses:2024sin,PHENIX:2024vjp,Csanad:2024hva} favor a more general description in the form of Lévy-stable shape~\cite{Csorgo:2003uv, Csanad:2007fr}, featuring a power-law tail. Several mechanisms are known to exhibit such tails~\cite{Csorgo:2004sr, Csorgo:2005it}, however, the most prominent process is Lévy-walk~\cite{Kincses:2024lnv}.

In several phenomenological studies~\cite{Kincses:2024lnv, Kincses:2025izu, Huang:2025edi, Porfy:2025sev, Molnar:2026vgi}, we can observe discrepancies between the source parameters obtained from theoretical models and those extracted from experimental data. One possible reason for such differences could stem from the fact that these models only roughly approximate the Coulomb interaction between emitted particles. As is known, the Coulomb interaction is long-ranged, hence, it may modify the large-distance behavior of the source and thereby affect the source shape parameter, the Lévy stability index $\alpha$, that governs the power-law tail of the spherically symmetric Lévy distribution, defined as:
\begin{equation}
    \mathcal{L}(\vec{r}; \alpha, R) = \frac{1}{(2\pi)^3} \int d^3 \vec{\zeta} e^{i\vec{\zeta}\cdot\vec{r}} e^{-\frac{1}{2}|\vec{\zeta}R|^\alpha},
\end{equation}
where $\vec{\zeta}$ is the integration variable, $\vec{r}$ is the variable of the distribution, $R$ is the Lévy scale parameter, and $\alpha$ is the Lévy-stability index ($0 < \alpha \leq 2$). If $\alpha = 1$, one gets back the Cauchy distribution, and for $\alpha=2$, one recovers the Gaussian distribution.

In the present work, we investigate how the Coulomb interaction influences the shape parameter $\alpha$ using custom-developed simulation programs written in C++ and Python. The C++ code integrates the relativistic equation of motion of the pions with the Leapfrog algorithm~\cite{Birdsall1985, Hairer2016}, starting from a generated source with a known stability index. The Python code constructs the pair-source distribution from the final phase-space coordinates similarly to Refs.~\cite{Kincses:2024lnv, Kincses:2025iaf, Kincses:2025izu, Huang:2025edi, Porfy:2025sev, Molnar:2026vgi, Kincses:2026tfa}, which is then fitted with a one-dimensional Lévy distribution~\cite{LevyCode} in the tail region. We then examine the ratio of the fitted to the generated $\alpha$, as well as their fitted slope, as a function of the number of particles, the fitting range of the pair-source distribution, and the initial source size and density.

The paper is organized as follows. Section~\ref{s:femto} briefly summarizes femtoscopy theory, Section~\ref{s:analysis} describes the simulation setup, forces and time propagation, Section~\ref{s:results} presents the results, and Section~\ref{s:summary} gives a summary.

\section{Femtoscopy}\label{s:femto}
In experimental measurements, the pair-source distribution $D$ cannot be measured directly, however, it can still be measured indirectly by measuring the correlation function $C(q, K)$. The pair-source distribution is defined as follows:
\begin{equation}
    D(r,K) = \int d^4\sigma \, S(\sigma + r/2,K)\,S(\sigma - r/2,K),
\end{equation}
where $r$ is the four-vector pair separation, $K$ is the pair average momentum, and $\sigma = (x_1 + x_2)/2$ is the pair center of mass. The pair-source distribution is the autoconvolution of the source $S(x,p)$ in the first variable. In the following, we use the three-vector of $r$ and $p$, and their derived variables; these are denoted with bold letters. 

The connection between $C(\boldsymbol{q})$ and $D(\boldsymbol{r})$ can be written as follows:
\begin{equation}
	C(\boldsymbol{q}) = \int d^3\boldsymbol{r} \, D(\boldsymbol{r})\,|\Psi^{(2)}_{\boldsymbol{q}}(\boldsymbol{r})|^2,
\end{equation}
where $\boldsymbol{r}$ is the pair separation, $\boldsymbol{q}$ is the momentum difference of the pair, and $\Psi^{(2)}_{\boldsymbol{q}}(\boldsymbol{r})$ is the pair-wavefunction at relative momentum $\boldsymbol{q}$. We furthermore note that the dominant variable of $C$ is $\boldsymbol{q}$, and of $D$ it is $\boldsymbol{r}$, and the dependence on pair average momentum $K$ is assumed to be weaker and usually appears through the $K$-dependence of the fitted parameters; thus, it is suppressed. 

\section{Analysis Details}\label{s:analysis}
To better understand the effect of Coulomb interaction in Lévy-distributed sources, we first need a Lévy-distributed set of particles. This can be done by using theoretical models, however, it is not trivial when and how one should extract these particles from said generators. Instead, we turn towards generating our cloud of particles, thus, we detail this in the following part.
\subsection{Kanter method}\label{ss:kantermethod}
We first generate a cloud of particles with a three-dimensional, spherically symmetric Lévy-stable distribution with $\alpha_0$ stability index. Since the densities of almost all $\alpha$-stable distributions have no known
closed form~\cite{Jacobs_2010}, they must be generated numerically. We sample a positive, totally skewed ($\beta=1$) stable variable $V$ with index $a=\alpha_0/2$ using the Kanter method~\cite{Kanter:1975}:
\begin{equation}
    V = \frac{\sin(a x)}{\left(\sin x\right)^{1/a}}
        \left( \frac{\sin[(1-a)x]}{y} \right)^{\frac{1-a}{a}},
\end{equation}
where $x$ is uniformly distributed on $(0,\pi)$, and $y$ has standard exponential density. The initial position of each particle is then
\begin{equation}
    \boldsymbol{\hat{r}} = r_0 \sqrt{V}\,\boldsymbol{G},
\end{equation}
where $\boldsymbol{G}$ is a vector of three independent standard normal variables and $r_0$ is the initial size of the source. As the same $V$ multiplies all three components, the direction is set by the isotropic $\boldsymbol{G}$, and the skewness of $V$ impacts only the radial distance. This approach then results in a distribution of $\boldsymbol{\hat{r}}$ that is spherically symmetric, and each of its Cartesian components follows a one-dimensional symmetric ($\beta=0$) stable distribution with $\alpha_0$. We have the position, thus, we only need the charge and momentum for each particle.

\subsection{Momentum and charge initialization}
The charge of particles was randomized with equal probability to -1 or 1, independently of their position and momentum. The momenta of each particle are generated independently of their positions, with their magnitude $p$ sampled from an exponential distribution. The direction of momentum vectors is chosen isotropically, with the azimuthal angle $\varphi$ uniformly distributed on $[0,2\pi)$ and
$\cos\vartheta$ uniformly distributed on $[-1,1)$, giving a uniform distribution on the unit sphere. The momentum vector takes the following form:
\begin{equation}
    \boldsymbol{p} = 
        \begin{bmatrix}
            \sin\vartheta\cos\varphi \\
            \sin\vartheta\sin\varphi \\
            \cos\vartheta
        \end{bmatrix}.
\end{equation}

\subsection{Debye screening}\label{ss:debye}
Debye screening accounts for the screening effect of the medium. To include it, the Coulomb potential must be replaced by the Yukawa potential,
\begin{equation}
    V(r) \propto \frac{1}{r}\, \exp(-r/\lambda_D),
\end{equation}
where $V(r)$ is the screened Coulomb potential between two particles in the expanding medium, $r$ is the distance between the two particles, and $\lambda_D$ is the Debye length. However, this is still not accurate enough, because as the medium expands, its temperature and particle density decrease. The Debye length is therefore time-dependent, and it must be written as
\begin{equation}
    \lambda_D(t) = \sqrt{\frac{T(t)}{4\pi \alpha_{\text{em}}\, n(t)}},
\end{equation}
where $T(t)$ is the temperature of the expanding medium, and $n(t)$ is the particle density at time $t$. Here $\alpha_{\text{em}}$ is the electromagnetic coupling constant, and the factor $4\pi$ arises from the three-dimensional symmetry. As time progresses, the Debye length increases, the particles move farther apart, and they decouple from the hot, expanding medium. At this stage, the Coulomb effect becomes negligible, and consequently the pair distribution function $D(r)$ stabilizes.

\subsection{Acting force}
We now have a cloud of charged particles, with $\alpha_0$ Lévy-distributed positions and $\boldsymbol{p}$ momenta, however, we have not yet added any force that would act on these particles. To describe the acting Coulomb force in short range, we use the Yukawa force, and to handle long-range interactions, we apply a time-dependent Debye screening $\lambda_D(t)$, mentioned in Sec.~\ref{ss:debye} . This force can be written as:
\begin{align}
    \mathbf{F}_{ij}^{\text{Coulomb}} &= \frac{e^2 q_i q_j}{r_{ij}^3}\\
    \mathbf{F}_{ij}^{\text{Yukawa}} &= \mathbf{F}_{ij}^{\text{Coulomb}} \cdot \left( 1 + \frac{r_{ij}}{\lambda_D(t)} \right) \cdot e^{-r_{ij}/\lambda_D(t)}, \mathbf{r}_{ij}
\end{align}
where $q$ is the charge of the given particle, $r_{ij}$ is the relative distance, the exponential part is responsible for the distance-dependent screening, and finally $\mathbf{r}_{ij}$ gives the direction of the interaction. To implement this, we first calculate the acceleration of each active particle $i$ as the sum of a mean-field term and Debye length-screened pairwise forces. 

To implement this, we first calculate the effect of the global mean field exerted by the expanding charged cloud on each particle. Next, we build (and refresh) a k-dimensional tree of the three-dimensional space, which allows us to exclude particles outside of the Debye length\footnote{This step allowed us to reduce the calculation time from $\mathcal{O}(N^2)$ to $\mathcal{O}(N \log N)$.}. In the third step, we do a neighbor search and sum the screened Yukawa forces. In this part, to avoid the singularity at small separations ($r \to 0$), the distance is softened,
\begin{equation}
    r_{ij} = \sqrt{|\boldsymbol{r}_i-\boldsymbol{r}_j|^2 + \varepsilon^2},
\end{equation}
where $\varepsilon$ is the softening length.

\subsection{Time evolution}
To numerically propagate the equation of motion, we used the Leapfrog integration with an adaptive time step $\Delta t$:
\begin{equation}
    \Delta t = 0.15 \cdot \sqrt{\frac{\varepsilon}{a_{\text{max}} + 10^{-6}}},
\end{equation}
where $a_{\text{max}}$ is the acceleration.

First, the momentum is calculated at half the time step ($\frac{\Delta t}{2}$) as
\begin{equation}
    \boldsymbol{p}\left(t+\tfrac{\Delta t}{2}\right) = \boldsymbol{p}(t) + \boldsymbol{F}(t)\frac{\Delta t}{2}.
\end{equation}
then, obtaining the half time step velocity using $p\left(t + \frac{\Delta t}{2}\right)$, we shift the particle with one time step:
\begin{equation}
    \boldsymbol{r}(t+\Delta t) = \boldsymbol{r}(t) + \frac{\boldsymbol{p}\left(t+\frac{\Delta t}{2}\right)}{\sqrt{m^2 + \boldsymbol{p}^2\left(t+\frac{\Delta t}{2}\right)/c^2}}\Delta t.
\end{equation}
The force is then recalculated at the new positions and time, and the momentum update is completed,
\begin{equation}
    \boldsymbol{p}(t+\Delta t) = \boldsymbol{p}\left(t+\frac{\Delta t}{2}\right) + \boldsymbol{F}(t+\Delta t)\frac{\Delta t}{2}.
\end{equation}
This method ensures that the total energy of the system stays bounded instead of accumulating over long times.

The default run time of the simulation is $t = 1000\text{ fm}/c$, however, it can be stopped earlier if certain conditions are met. Notably, if the momentum change of a given particle was less than a predefined $\delta p_\text{cut}$ three times in a row, the particle becomes inactive, and the propagation is stopped.

\section{Results}\label{s:results}
\begin{figure}[]
    \centering
    \includegraphics[width=1\textwidth]{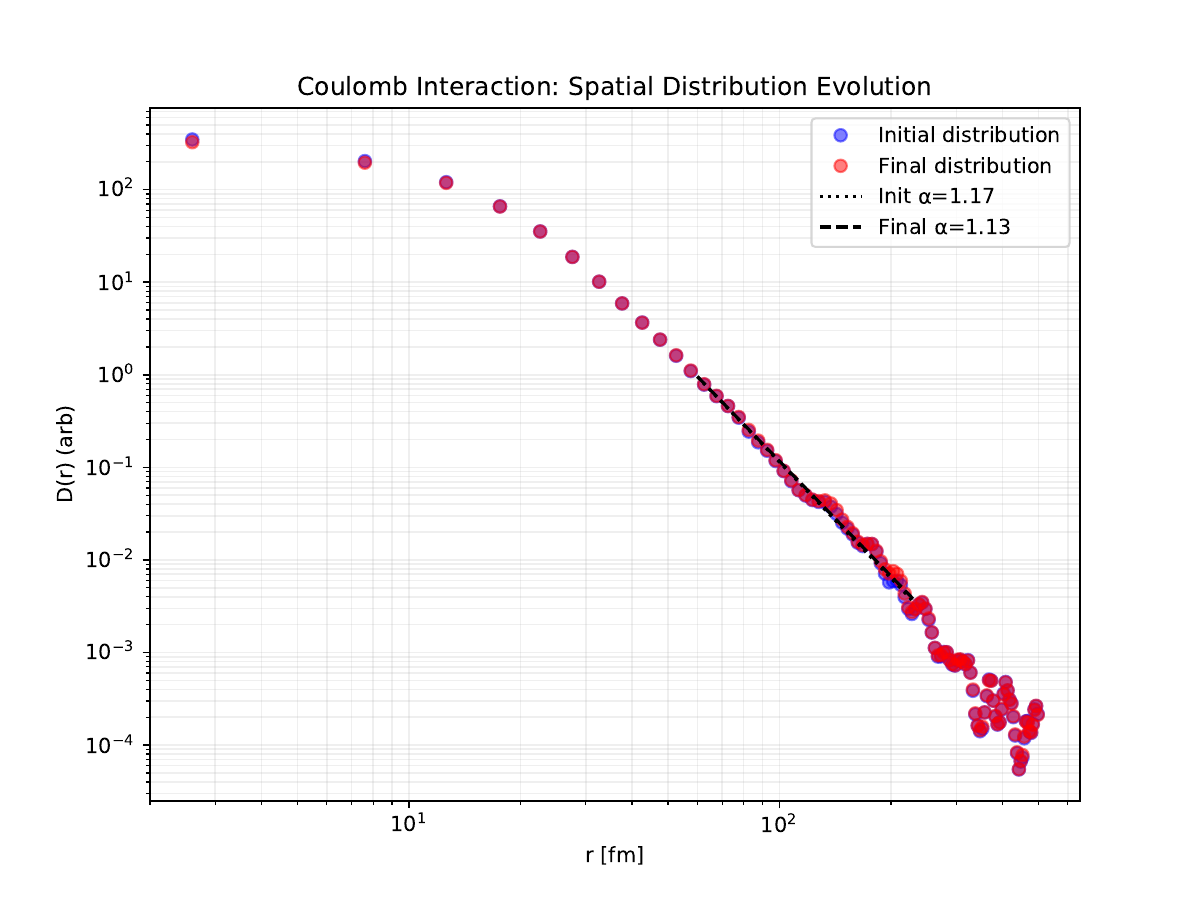}
    \caption{The pair-source distribution using the initial distribution (red full circles) and the final, finished simulation distribution (blue full circles). The fitted Lévy is also displayed with dotted and dashed lines.}
    \label{fig:Dr_comparison}
\end{figure}

As a first step, we compare the generated pair-source distribution $D(r)$ with the $D(r)$ obtained at the end of the simulation (Fig.~\ref{fig:Dr_comparison}). Both distributions were fitted with a Lévy distribution in the range $60 < r < 230$ fm. We obtained a small change in $\alpha$ ($\alpha_0 = 1.17$ and $\alpha_\text{final} = 1.13$), however, it is not prominent enough to draw a proper conclusion from a single simulation.

\begin{figure}[]
    \centering
    \includegraphics[width=1\textwidth]{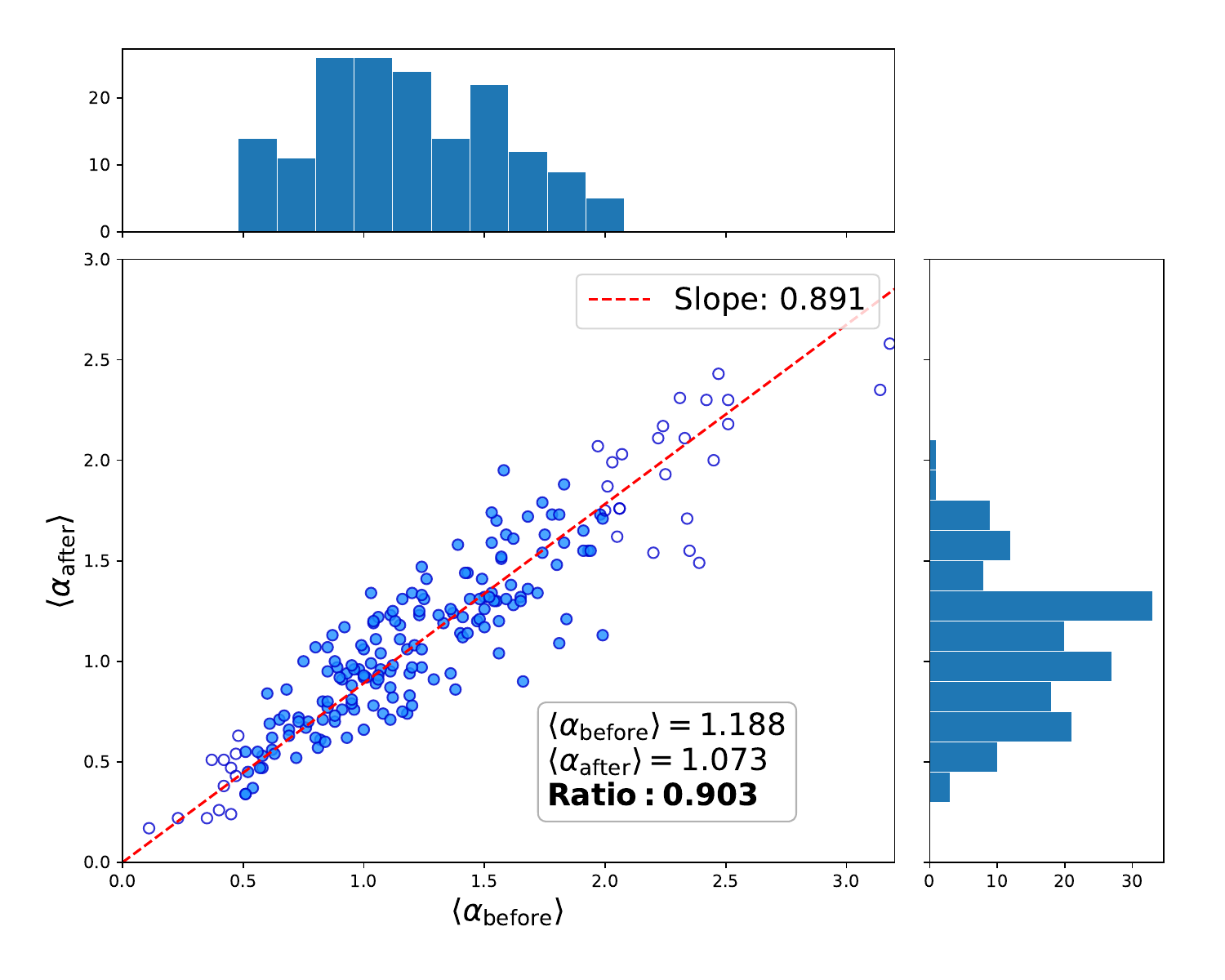}
    \caption{Scatter plot of several $D(r)$ fits, displaying $\alpha_0$ versus $\alpha_\text{final}$. The obtained ratio as well as the slope parameter is displayed.}
    \label{fig:alpha_comparison}
\end{figure}

To quantify the effect, we performed several simulations and plotted the initial $\alpha$ versus the extracted $\alpha$ (Fig.~\ref{fig:alpha_comparison}). In the following, we characterize the results by two quantities: the ratio $\alpha_{\text{final}}$/$\alpha_{0}$, and the slope $m$ of a linear fit $y = mx + b$ to the scatter plot. By obtaining the slope, we can study the relative change due to the Coulomb interaction, $m=1$ means no distortion. In the following part, we will discuss how the Coulomb interaction affects the ratio and the slope. Below, we systematically study how these two quantities depend on the particle number, the initial source size, the particle density, the fit range, and $\alpha_0$.

\subsection{Particle number}

\begin{figure}[]
    \centering
    \includegraphics[width=1\textwidth]{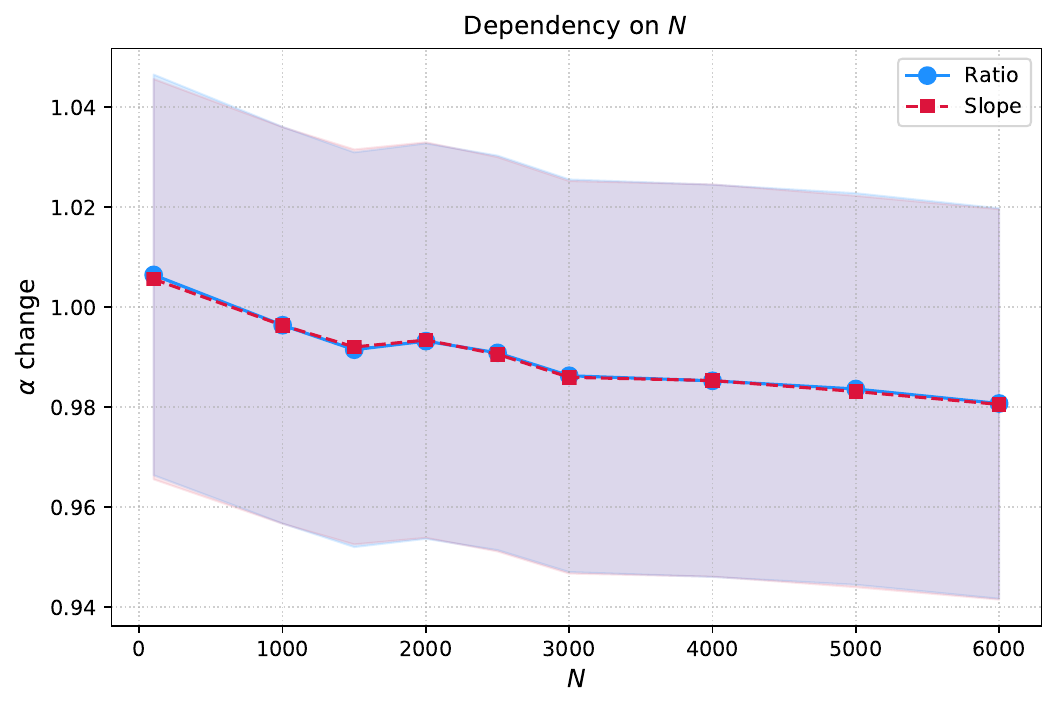}
    \caption{The obtained ratio (blue circle) and the slope parameter (red square) as a function of particle number $N$. The colored band shows the systematic uncertainty calculated from parameter fluctuations.}
    \label{fig:N_comparison}
\end{figure}

We first investigate the Lévy-index versus the particle number $N$. It is immediately visible in Fig.~\ref{fig:N_comparison} that the two observables are qualitatively the same and follow the same trend. In both cases, higher $N$ means a lower ratio and slope, signaling a stronger effect of Coulomb interaction. We observe a small change from about $1.01$ to $0.98$ in the studied range.

\subsection{Initial source size}
\begin{figure}[]
    \centering
    \includegraphics[width=1\textwidth]{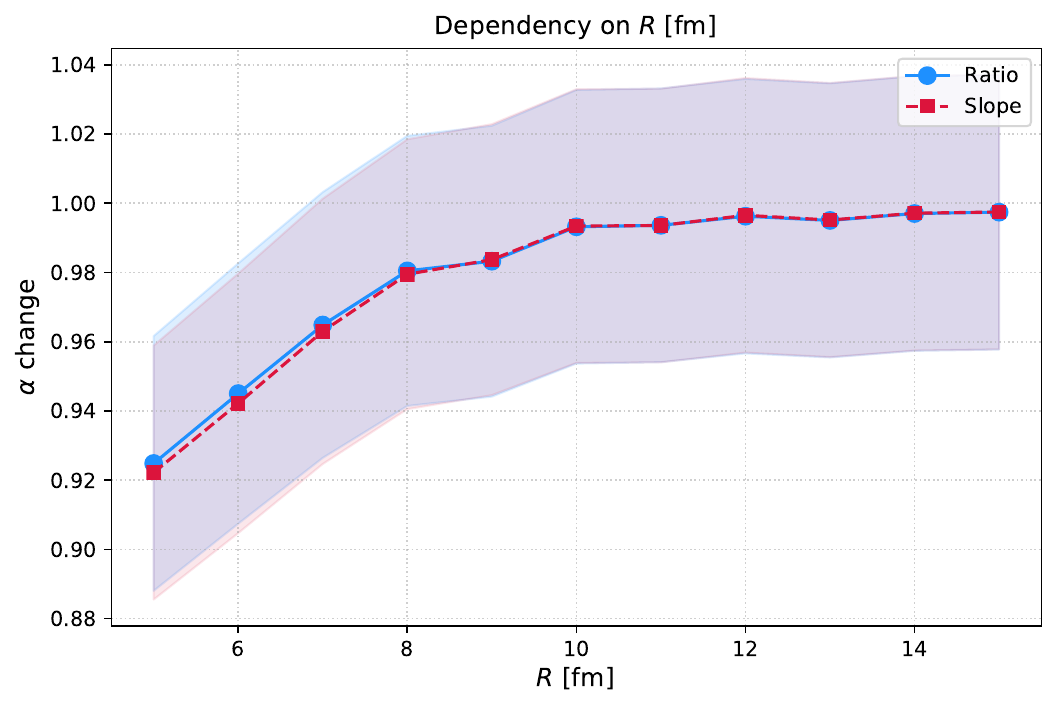}
    \caption{The obtained ratio (blue circle) and the slope parameter (red square) as a function of initial size $R_\text{init}$. The colored band shows the systematic uncertainty calculated from parameter fluctuations.}
    \label{fig:Rinit_comparison}
\end{figure}

Figure~\ref{fig:Rinit_comparison} shows that at lower source sizes, the difference between the two observables is more pronounced, while at larger initial sizes, the two are on top of each other. We observe that a larger initial size lowers the distortion effect of Coulomb. Furthermore, the change over the range is more visible, starting at about $0.92$ and rising to $0.99$ in the studied range.

\subsection{Particle density}
\begin{figure}[]
    \centering
    \includegraphics[width=1\textwidth]{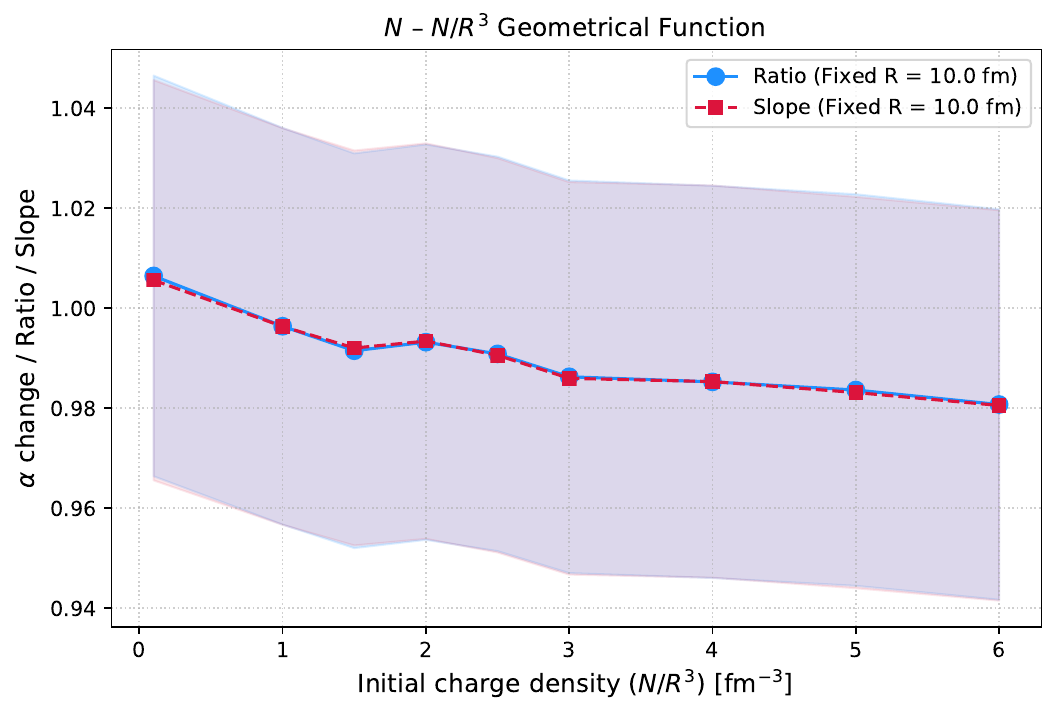}
    \includegraphics[width=1\textwidth]{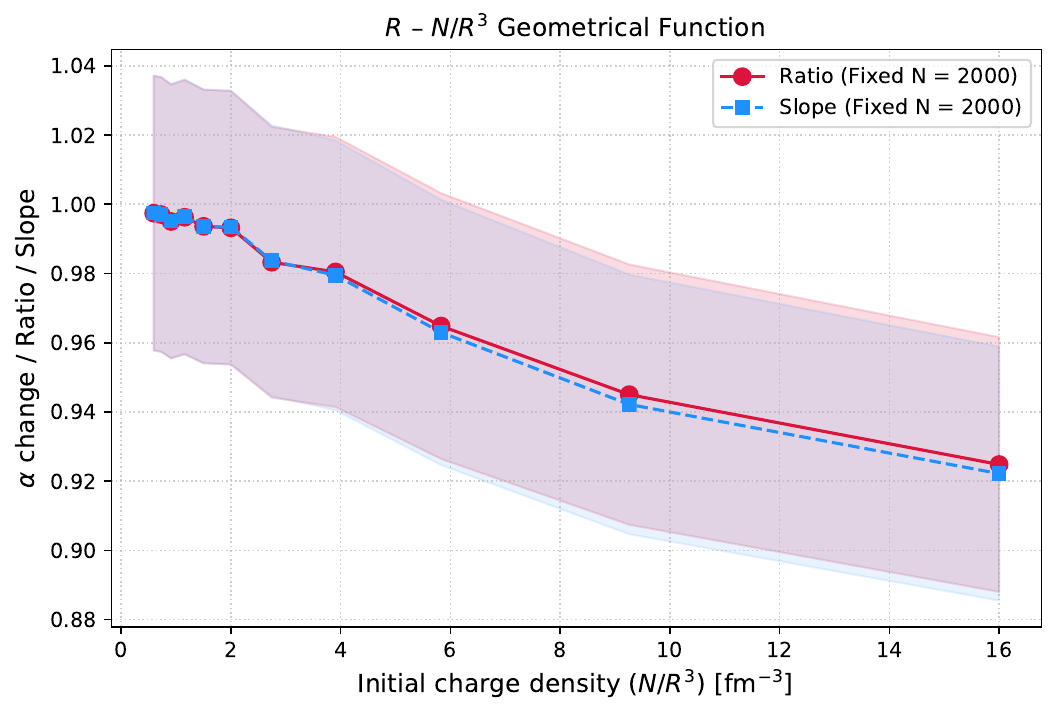}
    \caption{The obtained ratio (blue circle) and the slope parameter (red square) as a function of initial charge density $N / R^3 [\text{fm}^3]$. The colored band shows the systematic uncertainty calculated from parameter fluctuations. Top) while keeping $R = 10$ fm and bottom) while keeping $N = 2000$.}
    \label{fig:density_comparison}
\end{figure}

Next, we investigate the ratio and slope as a function of initial charge density ($N / R^3 [\text{fm}^3]$) of the cloud, visible in Fig.~\ref{fig:density_comparison}. We keep one parameter fixed at a time: first $R = 10$ fm, then $N = 2000$. The top plot of the figure (varying particle number) shows a slight decrease with charge density, whereas the bottom plot (varying initial size) shows a stronger decrease, signaling that the distortion is weaker on the number of particles.

\subsection{Initial $\alpha$}
\begin{figure}[]
    \centering
    \includegraphics[width=1\textwidth]{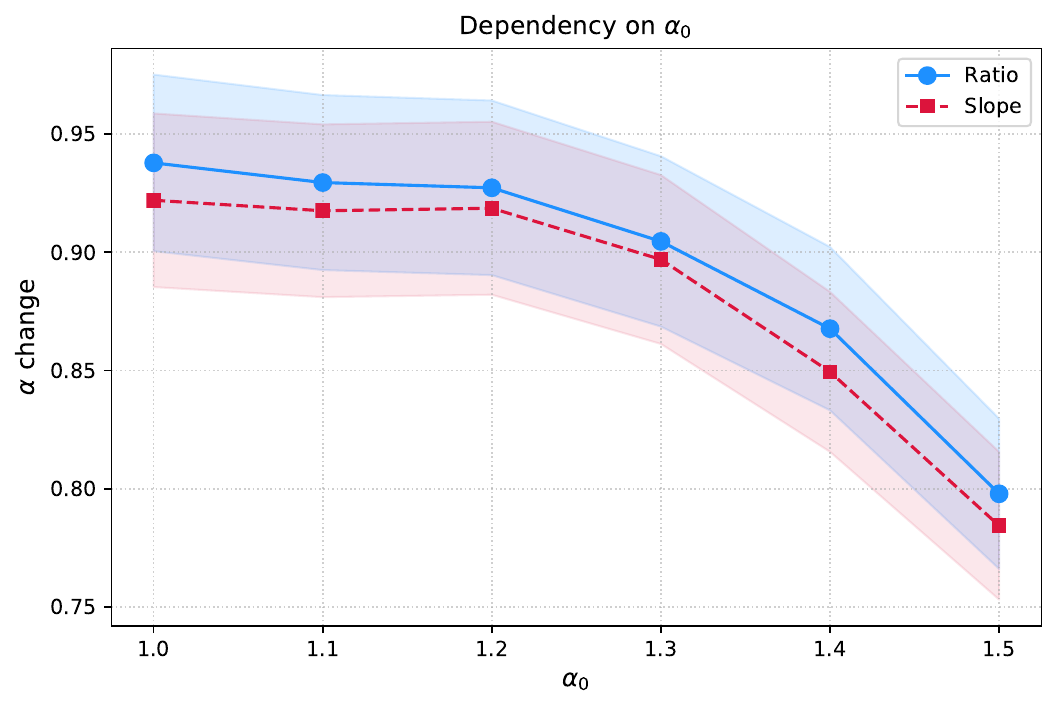}
    \caption{The obtained ratio (blue circle) and the slope parameter (red square) as a function of initial Lévy-index $\alpha_0$. The colored band shows the systematic uncertainty calculated from parameter fluctuations.}
    \label{fig:alpha0_comparison}
\end{figure}

Quite naturally, the initial Lévy-index also affects the two observables. We studied their dependence on $\alpha_0$ in Fig.~\ref{fig:alpha0_comparison}. Starting from Cauchy sources ($\alpha = 1$), we can observe a strong decrease of both ratio and slope as the value shifts towards $\alpha=1.5$. A possible explanation is that at fixed source size, a larger $\alpha_0$ concentrates more particles in the core, which strengthens the interaction, hence a stronger distortion. In addition, for larger $\alpha_0$, the tail of $D(r)$ falls off faster (e.g. Fig. 4 of Ref.~\cite{Kincses:2024lnv}), so the pairs might get pushed outward by the Coulomb repulsion, forming a larger relative contribution to the fitted region. 

\subsection{Fit range}
\begin{figure}[]
    \centering
    \includegraphics[width=1\textwidth]{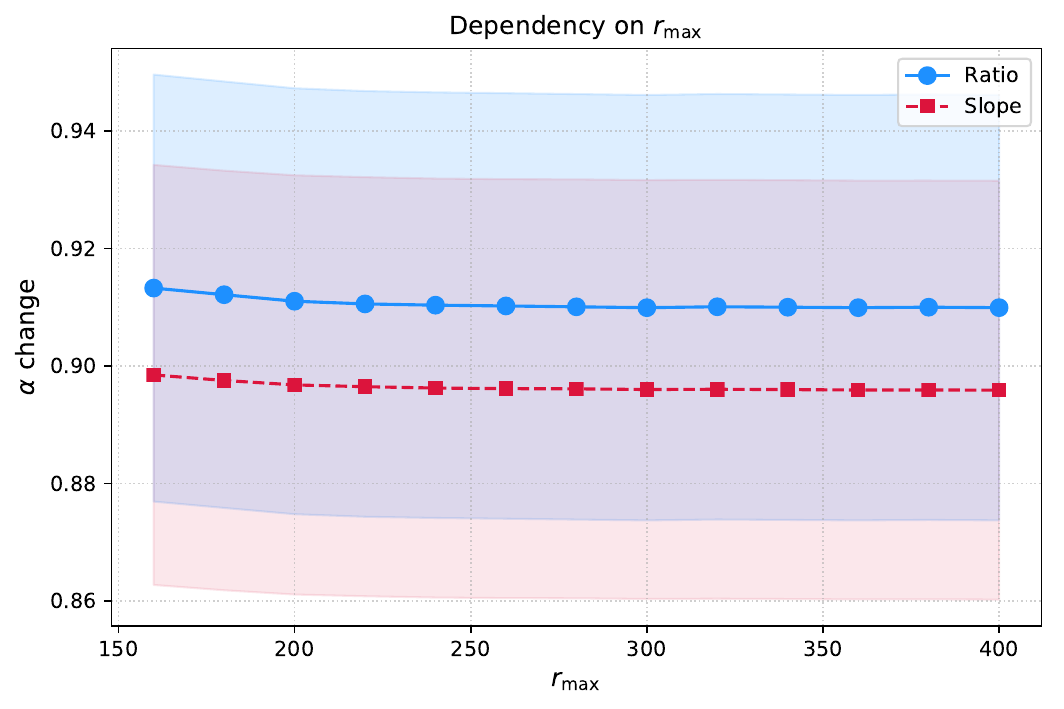}
    \caption{The obtained ratio (blue circle) and the slope parameter (red square) as a function of maximal fit range ($r_\text{max}$) on $D(r)$. The colored band shows the systematic uncertainty calculated from parameter fluctuations.}
    \label{fig:rmax_comparison}
\end{figure}

Finally, we investigated the dependence of the maximal fit range ($r_\text{max}$) on $D(r)$. In Fig.~\ref{fig:rmax_comparison}, we can see that both observables are very stable over the studied range.

\section{Conclusion}\label{s:summary}
We investigated the effect of Coulomb interaction on the Lévy stability index $\alpha$ of a particle-emitting source. We used a custom-developed simulation framework instead of the available models, where we generated three-dimensional, spherically symmetric Lévy-stable sources with a predefined $\alpha_0$ via the Kanter method. We then propagated the charged pions with the Leapfrog integration method, including time-dependent Debye screening, using Yukawa forces and constructed the pair-source distribution $D(r)$. After fitting $D(r)$ with a Lévy distribution, the distortion caused by the Coulomb interaction was characterized by the ratio $\alpha_\text{final}/\alpha_0$ and by the slope of the linear fit to the $\alpha_\text{final}$--$\alpha_0$ scatter.

We found that the Coulomb interaction generally lowers $\alpha_\text{final}$, thus making the tail heavier. This effect was observed to grow with increasing particle number, and weaken with increasing initial source size. Studying the initial charge density showed that changing the source size has a stronger impact compared to changing the particle number at the same density. The distortion also increases strongly with increasing the initial Lévy index from $\alpha_0 = 1$ to $1.5$, which might be attributed to the higher initial density and the faster-falling tail of sources. Finally, both observables were found to be stable with respect to the upper limit of the fit range.

Our methodology, as discussed above, is based on fitting the power-law tail of the late-time pair-separation distribution, which indicates a decrease in $\alpha$ after the Coulomb evolution. This result may be influenced by methodological effects: the late-time positions contain free streaming unrelated to the interaction, and we effectively fit only the power-law tail. In a refined setup, one could define emission points by back-extrapolating the asymptotic straight-line trajectories, use a regulator to replace the fixed short-distance cutoff, and fit the observed distributions with a correctly scaled three-dimensional L\'evy density. Hence our results shall be regarded as qualitative, and more refined calculations should follow to explore this effect in detail.

\section*{Acknowledgments}
This work was supported by the NKFIH grant NKKP ADVANCED 152097. Furthermore, Barnabás Pórfy was supported by the DKÖP-23 Doctoral Excellence Program of the Ministry for Culture and Innovation and by the EKÖP-26 University Research Scholarship Program of the Ministry of Science and Technology from the source of the National Research, Development and Innovation Fund. Kristóf Csiki was supported by the ELTE Talentum Program.

\bibliographystyle{ws-ijmpa.bst}
\bibliography{wpcf}

\begin{thebibliography}{10}
\bibitem{LevyCode}
M.~Csanad, { Code for calculating angular average and projections of a full 3d lévy-stable distribution and fit it to data}.
\end{thebibliography}
\end{document}